# Electric Field-Tuned Topological Phase Transition in Ultra-Thin Na$_3$Bi – Towards a Topological Transistor

James L. Collins[1,2,3], Anton Tadich[3,4], Weikang Wu[5], Lidia C. Gomes[6,7], Joao N. B. Rodrigues[6,8], Chang Liu[1,2,3], Jack Hellerstedt[1,2], Hyejin Ryu[9], Shujie Tang[9], Sung-Kwan Mo[9], Shaffique Adam[6,10], Shengyuan A. Yang[5], Michael. S. Fuhrer[1,2,3], Mark T. Edmonds[1,2,3*]

1. School of Physics and Astronomy, Monash University, Clayton VIC 3800, Australia
2. Monash Centre for Atomically Thin Materials, Monash University, Clayton VIC 3800 Australia
3. Centre for Future Low Energy Electronics Technologies, Monash University, Clayton VIC 3800 Australia
4. Australian Synchrotron, 800 Blackburn Road, Clayton, VIC 3168, Australia
5. Research Laboratory for Quantum Materials & EPD Pillar, Singapore University of Technology and Design, Singapore 487372, Singapore
6. Department of Physics and Centre for Advanced 2D Materials, National University of Singapore, 117551, Singapore
7. National Centre for Supercomputing Applications, University of Illinois at Urbana-Champaign
8. Institute for Condensed Matter Theory and Department of Physics, University of Illinois at Urbana-Champaign
9. Advanced Light Source, Lawrence Berkeley National Laboratory, Berkeley, CA 94720, USA
10. Yale-NUS College, 6 College Avenue East, 138614, Singapore

**\* Corresponding author - mark.edmonds@monash.edu**

**Abstract –**

**The electric field induced quantum phase transition from topological to conventional insulator has been proposed as the basis of a topological field effect transistor [1-4]. In this scheme an electric field can switch 'on' the ballistic flow of charge and spin along dissipationless edges of the two-dimensional (2D) quantum spin Hall insulator [5-9], and when 'off' is a conventional insulator with no conductive channels. Such a topological transistor is promising for low-energy logic circuits [4], which would necessitate electric field-switched materials with conventional and topological bandgaps much greater than room temperature, significantly greater than proposed to date [6-8]. Topological Dirac semimetals (TDS) are promising systems in which to look for topological field-effect switching, as they lie at the boundary between conventional and topological phases [3,10-16]. Here we use scanning probe microscopy/spectroscopy (STM/STS) and angle-resolved photoelectron spectroscopy (ARPES) to show that mono- and bilayer films of TDS Na$_3$Bi [3,17] are 2D topological insulators with bulk bandgaps >400 meV in the absence of electric field. Upon application of electric field by doping with potassium or by close approach of the STM tip, the bandgap can be completely closed then re-opened with conventional gap greater than 100 meV. The large bandgaps in both the conventional and quantum spin Hall phases, much**

**greater than the thermal energy $kT = 25$ meV at room temperature, suggest that ultrathin Na$_3$Bi is suitable for room temperature topological transistor operation.**

Two-dimensional Quantum Spin Hall (QSH) insulators are characterized by an insulating interior with bulk bandgap $E_g$, and topologically protected conducting edge channels that are robust to backscattering by non-magnetic disorder. The QSH effect was first realized in HgTe quantum wells [5] where the small $E_g$ prevents device applications above cryogenic temperatures. This has led to efforts to find new materials with $E_g \gg 25$ meV (thermal energy at room temperature) for room temperature topological electronic devices. Recent reports of QSH insulators bismuthene on SiC ($E_g \sim 0.8$ eV) [6] and monolayer 1T′-WTe$_2$ ($E_g \sim 50$ meV) [7] are promising, with the quantum spin Hall effect measured in monolayer WTe$_2$ up to 100K [8,9]. However, a predicted electric-field effect in WTe$_2$ has not yet been reported, and a substantial field-effect in atomically two-dimensional bismuthene is unlikely due to the completely in-plane structure that lacks inversion symmetry meaning any Stark effect would most likely be small.

Ultrathin films of topological Dirac semimetals (TDS) are a promising material class to realise the electric-field tuned topological phase transition, with such a transition predicted in few-layer films of TDS Na$_3$Bi and Cd$_3$As$_2$ [3]. Bulk TDSs are zero-bandgap semimetals with a linear band dispersion in all three dimensions around pairs of Dirac points [10-13], whilst few-layer TDSs are predicted to be non-trivial insulators with bulk bandgaps up to ~300 meV for monolayer Na$_3$Bi [17]. However, experiments on few-layer TDSs are presently lacking [14-16]. To unambiguously demonstrate electric field control over the magnitude, electric field-dependence, and topological nature of the bandgap in ultrathin Na$_3$Bi, we employ two independent experimental techniques. First, we utilize ARPES to directly measure the electronic bandstructure and its modification as a result of doping the surface with potassium to generate an electric field. Second, we use scanning tunnelling spectroscopy (STS), which measures the local density of states (DOS) as a function of energy, to directly probe the energy gap while varying the tip-sample separation and consequently the induced electric field caused by the potential difference between tip and sample. STS also resolves the topological edge state in Na$_3$Bi at low electric field, demonstrating the topological nature of this phase.

The unit cell of Na$_3$Bi contains two stacked triple layers in the z direction, comprising Na and Bi atoms that form a honeycomb lattice, with interleaved Na atoms as shown in the crystal structures of Fig. 1(a)-(b). One triple layer and two stacked triple layers correspond to monolayer (ML) and bilayer (BL) Na$_3$Bi respectively as illustrated in Fig. 1(b). Figure 1(c) shows the Brillouin zone of 2D Na$_3$Bi. In Fig. 1(d) STM on few-layer Na$_3$Bi(001) epitaxial films grown via molecular beam epitaxy (MBE) on Si(111) (see methods and SM1 for details) reveals coexisting regions of ML and BL Na$_3$Bi islands that are atomically flat and up to 40 nm in size, along with small areas of bare substrate. Monolayer regions are identified by an additional 0.22 nm distance to the underlying substrate, due to interfacial spacing or structural relaxation which has been observed previously in other atomically thin materials [18]. Figure 1(e) shows the overall band structure of few-layer Na$_3$Bi films along the M - Γ -K surface directions measured with ARPES at $hv = 48$ eV. Figure 1(f) shows the second derivative of the spectra in order to enhance low intensity features. This has been overlaid with density functional theory (DFT) calculations for ML (blue) and BL (green) Na$_3$Bi showing good agreement, consistent with the STM topography which shows coexisting ML and BL regions. Photon-energy dependent ARPES (see SM2) demonstrates that the film is electronically 2D, with no dispersion in $k_z$ unlike its bulk or 15 nm thin-film counterparts [10,15]. Depth dependent X-ray photoelectron spectroscopy (XPS) (See SM1), revealed no additional components observed in either the Si2p core level corresponding to the Si(111) substrate or the Na and Bi core levels of $Na_3Bi$, verifying $Na_3Bi$ is free-standing on Si(111).

We first measure the size of the bulk bandgap for ML and BL regions of $Na_3Bi$ by probing the electronic structure with STS, which measures the differential conductance d$I$/d$V$ as a function of sample bias $V$, and is proportional to the local density of states. Figure 2(a) shows typical d$I$/d$V$ spectra for ML (red) and BL (black) with bandgaps corresponding to 0.44 ± 0.01 eV and 0.41 ± 0.01 eV respectively (see SM3 for details on extracting bandgap values). All STS curves in Fig. 2(a) were taken more than 5 nm inwards from step edges. Figure 2(b) plots the experimental bandgap (blue squares) in comparison to DFT calculated values for pristine $Na_3Bi$ (black circles) and $Na_3Bi$ layers that contain an Na(2) surface vacancy (red circles) (see Methods for details, with associated bandstructures found in SM4). The ~0.15 eV discrepancy between experiment and DFT for ML is likely a result of the underestimation of electronic bandgap well known to occur in DFT calculations [19]. The minimal change in bandgap from ML to BL observed experimentally is well explained by the DFT calculations that include Na(2) surface vacancies; this vacancy gives rise to a delocalized resonance feature and enhancement of the electronic bandgap [20], resulting in only a small layer-dependent evolution in bandgap.

To verify the prediction that ML and BL $Na_3Bi$ are large bandgap QSH insulators (see Fig. S4) we probe the step edge of these islands to the underlying Si(111) substrate to look for the conductive edge state signature of a QSH insulator (STM topography shown in Fig. 2(c)). Figure 2(d) shows local d$I$/d$V$ spectra taken on BL $Na_3Bi$ far away from the edge (black curve) and at the edge (blue curve). In contrast to the gap in the bulk, the d$I$/d$V$ at the edge is quite different, with states filling the bulk gap along with a characteristic dip at 0 mV bias. Similar features observed in other QSH insulators 1T'-$WTe_2$ [7] and bismuthene [6] have been attributed to one-dimensional (1D) non-trivial edge states and the emergence of a Luttinger liquid [21]. Figure 2(e) probes d$I$/d$V$ as a function of distance away from the edge, demonstrating the extended nature of the edge state feature, with Fig. 2(f) showing that the dI/dV signal within the bulk bandgap moving away from the edge follows the expected exponential decay for a 1D topological non-trivial state [6].

With ML and BL $Na_3Bi$ verified as large bandgap QSH insulators, we now examine the role of electric field on modifying the size and nature of the bandgap. First, we utilize ARPES to measure the bandstructure after doping the surface with K to generate an electric field. Details on calculating the displacement field are in SM5. Figure 3(a)-(d) shows the bandstructure along Γ - K for values of the electric field of 0.0, 0.72, 1.44 and 2.18 Vnm$^{-1}$ respectively, with the green and blue dots reflecting the extracted maxima from energy distribution curve (EDC) and momentum distribution curve (MDC) analysis (see SM5 for details). The right panel in (a)-(d) represents a model of a 2D gapped Dirac system (see SM5). In Fig. 3(a) only the hole band is observable, which away from Γ displays a linear band dispersion (hole Fermi velocity of $v_F$ ~ 3x10$^5$ ms$^{-1}$). The band dispersion near Γ displays the clear cusp of a band edge indicating a gapped system, with 140 meV separation between the valence band edge and the Fermi energy $E_F$. The effect of K dosing in Figs. 3(b)-(d) is to $n$-type dope the sample and consequently increase the displacement field. At a displacement field of 0.7 Vnm$^{-1}$ the separation from the valence band edge to $E_F$ has increased to ~257 meV. The bandgap must be at least this amount, consistent with STS, though we cannot determine its exact magnitude since the conduction band lies above $E_F$ (although it can be estimated see Fig. S7(a)). Upon increasing the displacement field, a Dirac-like electron band emerges (electron Fermi velocity of $v_F$ ~ 10$^6$ ms$^{-1}$). At 1.4 Vnm$^{-1}$ the extracted gap between the two band edges is ~100 meV, and reduces to ~90 meV at 2.2 Vnm$^{-1}$ (see Supplementary Materials S5 for calculation and comparison with STS values). Whilst a significant reduction in bandgap with displacement field occurs, due to the finite energy width of the bands (~100 meV) we cannot say definitively whether the gap is fully closed or even reopened again.

To elucidate the effect of an electric field on the electronic structure more clearly we turn back to measurements made with STM. Here, the tip-sample separation is now varied in order to tune the electric field due to the electrostatic potential difference between the metallic tip and $Na_3Bi$ as illustrated schematically in Fig. 4(a). The electrostatic potential difference is dominated by the difference in work function between the tip and the sample, being approximately 1.2 eV(see SM6 for calculation). Changes in the bandgap can then be measured in the d$I$/d$V$ spectra as a function of tip-sample separation and converted to electric field as shown in Fig. 4(b) (details of tip-sample distance calculation and electric field are found in SM6). Figure 4(b) shows four distinct normalized d$I$/d$V$ spectra taken on BL $Na_3Bi$ (see SM6 for similar spectra taken on ML $Na_3Bi$) at tip-sample separation (electric field) of 1.3 nm (0.9 Vnm$^{-1}$) (black curve), 1.1 nm (1.1 Vnm$^{-1}$) (blue curve), 0.8 nm (1.5 Vnm$^{-1}$) and 0.5 nm (2.4 Vnm$^{-1}$). A large modulation occurs upon increasing the electric field strength, with the bandgap reducing from ~400 meV to ~200 meV between 0.9 Vnm$^{-1}$ and 1.1 Vnm$^{-1}$, followed by the characteristic V-shape of a Dirac-semimetal at 1.5 Vnm$^{-1}$ indicating the bandgap is completely closed. At electric fields greater than 1.5 Vnm$^{-1}$ the bandgap reopens and reaches ~100 meV at 2.4 Vnm$^{-1}$. Figure 4(c) plots the bandgap as a function of electric field for ML and BL $Na_3Bi$, with both exhibiting a similar critical field where the bandgap is closed and then reopened into the trivial/conventional regime with increasing electric field.

By combining ARPES and STS, we have demonstrated that monolayer and bilayer $Na_3Bi$ are quantum spin Hall insulators with bulk bandgaps above 400 meV, offering the potential to support dissipationless transport of charge at room temperature. An electric-field tunes the phase from topological to conventional insulator with bandgap of ~100 meV. $Na_3Bi$ is chemically inert in contact with silicon, and the electric fields required to induce the topological phase transition are typically below the breakdown field of conventional dielectrics. These aspects make ultrathin $Na_3Bi$ a promising platform for realising new forms of electronic switches based on topological transistors for low-energy logic circuits. In addition the bandgap modulation of more than 400 meV is far larger than has been achieved in atomically thin semiconductors such as bi-layer graphene [22,23] and phosphorene [24], and may be useful in optoelectronic applications [25] in the mid-infrared.

## Methods:

### Growth of Ultra-Thin $Na_3Bi$ on Si(111)

Ultra-thin $Na_3Bi$ thin films were grown in ultra-high vacuum (UHV) molecular beam epitaxy (MBE) chambers and then immediately transferred after the growth to the interconnected measurement chamber (i.e. Createc LT-STM at Monash University, Scienta R-4000 analyser at Advanced Light Source and SPEC Phoibos 150 at Australian Synchrotron). To prepare an atomically flat substrate, a p-type Si(111) wafter was flash annealed to 1250°C in order to achieve a (7 x 7) surface reconstruction (this was achieved via direct current heating at Monash University and the Australian Synchrotron, and via electron bombardment heating at the Advanced Light Source).

For $Na_3Bi$ film growth effusion cells were used to simultaneously evaporate elemental Bi (99.999%) in an overflux of Na (99.95%) with a Bi:Na flux ratio not less than 1:10, calibrated by quartz microbalance. During growth the substrate temperature was between 300°C-330°C for successful crystallization. At the end of the growth the sample was left at the growth temperature for 10 min in an Na overflux to improve the film quality before cooling to room temperature. Reflection high energy electron diffraction (RHEED) and low energy electron diffraction (LEED) were used to confirm the (001) single-crystal epitaxial growth across a large area (see Supplementary Materials S1).

### STM/STS Measurements:

STM/STS measurements were performed in a Createc LT-STM operating at 5K. A PtIr tip was prepared and calibrated using an Au(111) single crystal and confirming the presence of the Shockley surface state at -0.5V and flat LDOS near the Fermi level before all measurements. After measurements on few-layer $Na_3Bi$ were completed the Au(111) sample was re-measured to confirm the tip had not significantly changed and still exhibited flat local density of states (LDOS) near the Fermi level. STM differential conductance measurements ($dI/dV$) were measured using a 5 mV AC excitation voltage (673 Hz) that was added to the tunnelling bias. Differential conductance measurements were made under open feedback conditions with the tip in a fixed position above the surface. Data was prepared and analysed using MATLAB and IGOR Pro.

### Angle resolved photoemission spectroscopy (ARPES) measurements:

ARPES measurements were performed at Beamline 10.0.1 of the Advanced Light Source (ALS) at Lawrence Berkeley National Laboratory, USA. Data was taken using a Scienta R4000 analyzer at 20 K sample temperature. The total energy resolution was 20-25 meV depending on the beamline slit widths and analyser settings, and the angular resolution was 0.2°. This resulted in an overall momentum resolution of ~0.01 Å$^{-1}$ for photoelectron kinetic energies measured, with the majority of the measurements performed at $hv = $ 48 eV and 55 eV. Doping of the $Na_3Bi$ films was performed via in-situ K dosing from a SAES getter source in UHV. Dosing was performed at 20K to avoid K clustering.

### XPS Measurements:

XPS measurements were performed at the Soft X-ray Beamline of the Australian Synchrotron using a SPECS Phoibos-150 spectrometer. The Bi 5d and Na 2p of the $Na_3Bi$, as well as the Si 2p of the Si(111) substrate were measured at photon energies of 100 eV, 350 eV, 800 eV and 1487 eV. This ensured surface sensitivity for the low photon energy scans at 100 eV, with the higher photon energies used to characterize the depth dependence of the core levels, in particular whether there was any chemical bonding between the $Na_3Bi$ film and the Si(111) substrate. The binding energy scale of all spectra are referenced to the Fermi energy ($E_F$), determined using either the Fermi edge or 4f core level of an Au reference foil in electrical contact with the sample. Core level spectra were analysed using a Shirley background subtraction.

**Density Functional Theory (DFT) Calculations:**

First-principles calculations based on density-functional theory (DFT) are used to obtain electronic band structures of monolayer and bilayer $Na_3Bi$, with and without Na(2) vacancies. This was achieved using the projector augmented wave (PAW) method [26] with calculations implemented in Quantum ESPRESSO code and the Vienna ab initio Simulation Package (VASP). The generalized gradient approximation (GGA) using the PBE functional [27] for the exchange-correlation potential were adopted. The plane-wave cutoff energy was set to be 400 eV. The Brillouin zone sampling was performed by using k grids with a spacing of $2\pi \times 0.02$ Å$^{-1}$ within a $\Gamma$-centered sampling scheme. For all calculations, the energy and force convergence criteria were set to be $10^{-5}$ eV and $10^{-2}$ eV/Å, respectively. For the $Na_3Bi$ layers, we used a vacuum region of thickness larger than 15 Å thickness to eliminate the artificial interaction between the periodic images. The edge states were studied by constructing the maximally localized Wannier functions [28] and by using the iterative Green function method [29] as implemented in the WannierTools package [30].

**Author Contributions**

M. T. E., J. L. C., and M. S. F. devised the STM experiments. M. T. E. devised the ARPES and XPS experiments. M. T. E. and J. L. C. performed the MBE growth and STM/STS measurements at Monash University. J. H. assisted with the experimental setup at Monash University. M. T. E., J. L. C. and A. T. performed the MBE growth and ARPES measurements at Advanced Light Source with the support from H. R., S. T. and S. K. M.. The MBE growth and XPS measurements at the Australian Synchrotron were performed by M. T. E., J. L. C. A. T., J. H., C. L. The DFT calculations were performed by L. C. G, J. N. B. R., W. W. and S. Y.. S. A. assisted with the theoretical interpretation of the data. J. L. C., M. T. E., and M. S. F. composed the manuscript. All authors read and contributed feedback to the manuscript.

**Acknowledgements**

M. T. E. is supported by ARC DECRA fellowship DE160101157. M. T. E., J. L. C., C. L, and M. S. F. acknowledge funding support from CE170100039. J. L. C., J. H and M. S. F. are supported by M. S. F.'s ARC Laureate Fellowship (FL120100038). M. T. E. and A. T. acknowledge travel funding provided by the International Synchrotron Access Program (ISAP) managed by the Australian Synchrotron, part of ANSTO, and funded by the Australian Government. M. T. E. acknowledges funding from the Monash Centre for Atomically Thin Materials Research and Equipment Scheme. S. Y. and W. W. acknowledge funding from Singapore MOE AcRF Tier 2 (Grant No. MOE2015-T2-2-144). This research used resources of the Advanced Light Source, which is a DOE Office of Science User Facility under contract no. DE-AC02-05CH11231. Part of this research was undertaken on the soft X-ray beamline at the Australian Synchrotron, part of ANSTO.

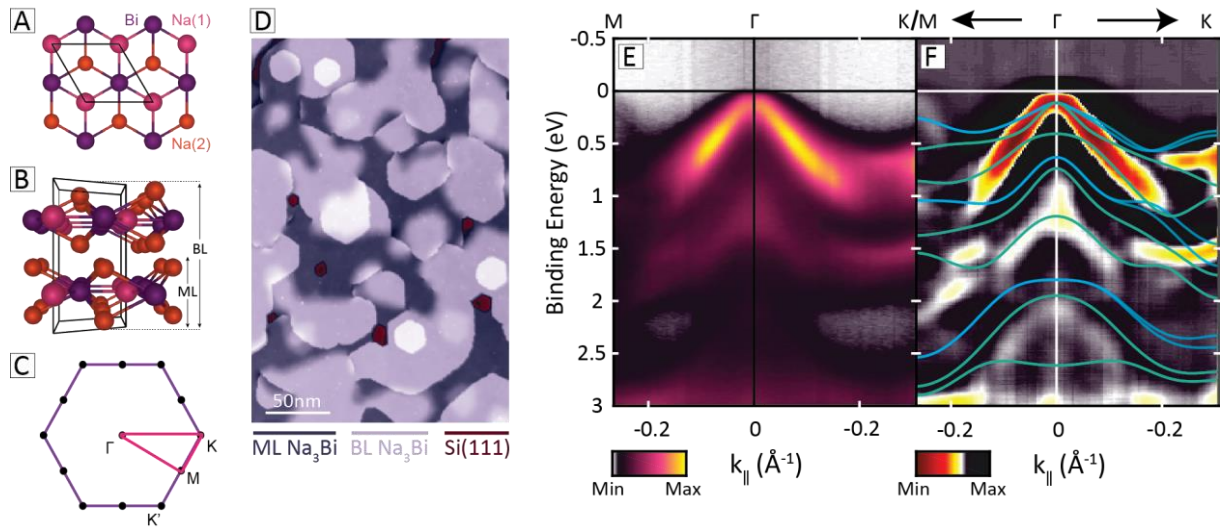

**Figure 1. Characterization of epitaxial few-layer Na₃Bi and overall electronic structure from ARPES.** (a) Top and (b) side view of the Na₃Bi unit cell. A monolayer represents a single triple-layer and bilayer represents two triple-layers. (c) Brillouin zone of 2D Na₃Bi. (d) Large area (300 nm x 200 nm) topographic STM image (bias voltage $V$ = -2.5 V and tunnel current $I$ = 100 pA) of few layer Na₃Bi on Si(111). Scale bar is 50 nm. The different regions are colour coded with ML (dark grey), BL (light grey) and Si(111) (red). (e) Overall band structure measured along the M-Γ-K directions measured with ARPES at $hv$=48 eV. (f) Second derivative spectra of (e) in order to enhance low intensity features. The overlaid blue and green curves are calculated DFT band structures for mono- and bi-layer Na₃Bi respectively.

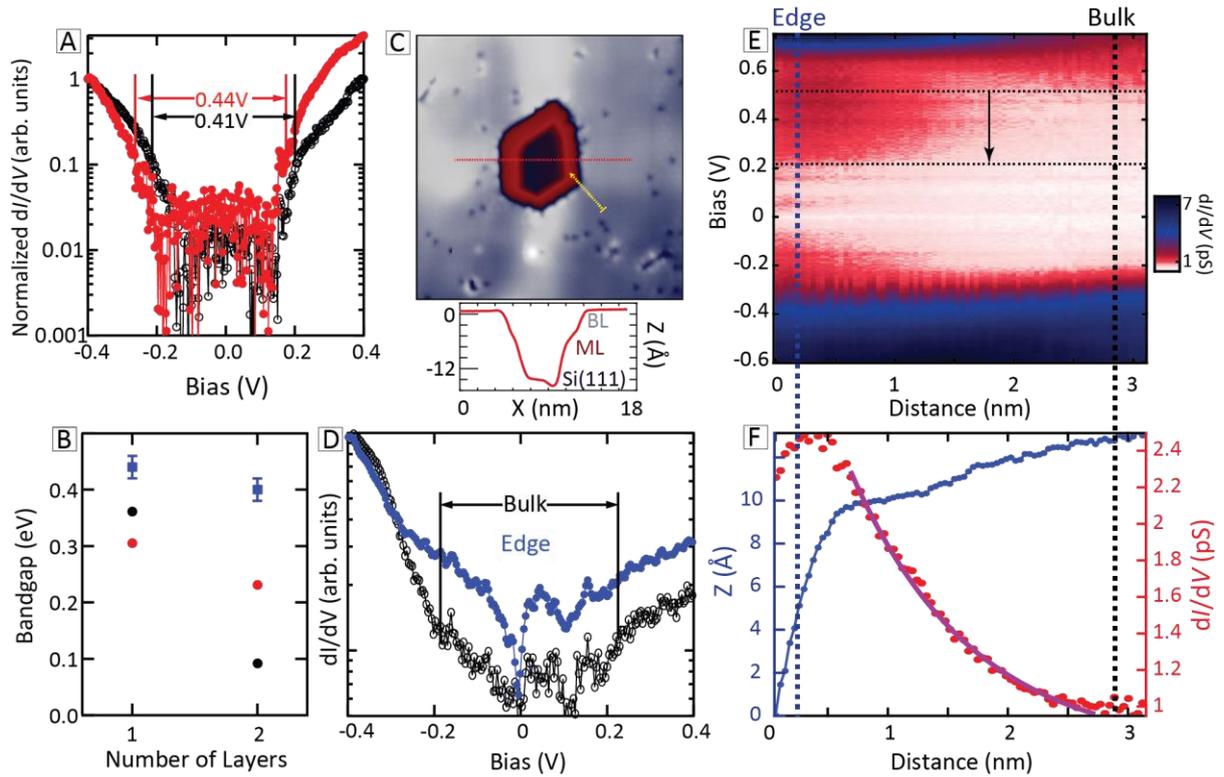

**Figure 2. Bandgap in mono- and bilayer Na$_3$Bi and edge state behaviour.** (a) Normalized d$I$/d$V$ spectra displayed on a logarithmic plot corresponding to ML (red) and BL (black) Na$_3$Bi. The conduction and valence band edges are reflected by the sharp onset of d$I$/d$V$ intensity. Details are found in SM3. (b) Evolution of the bandgap as a function of layer thickness determined from STS experiment (blue squares), DFT calculations on pristine Na$_3$Bi (black circles) and with an Na(2) vacancy (red circles). (c) STM topography of a region of bi-layer Na$_3$Bi (grey), monolayer Na$_3$Bi (red) and the underlying Si(111) substrate (black). The red line represents the step-height profile, and the yellow line reflects the region over which the dI/dV measurements were performed in (e). (d) d$I$/d$V$ spectra taken near the edge of BL Na$_3$Bi to the Si(111) substrate (blue) and in the bulk (black). (e) d$I$/d$V$ colour map taken at and then moving away from the step edge where the dashed vertical lines reflect the spectra shown in (d) and the horizontal lines represent the averaged dI/dV signal region that is shown in (f). (f) Shows the corresponding intensity profile of d$I$/d$V$ in the bulk gap showing the exponential decay away from the step edge, where the blue trace represents the topography height profile and red trace represents the dI/dV intensity within the horizontal dashed region of (e).

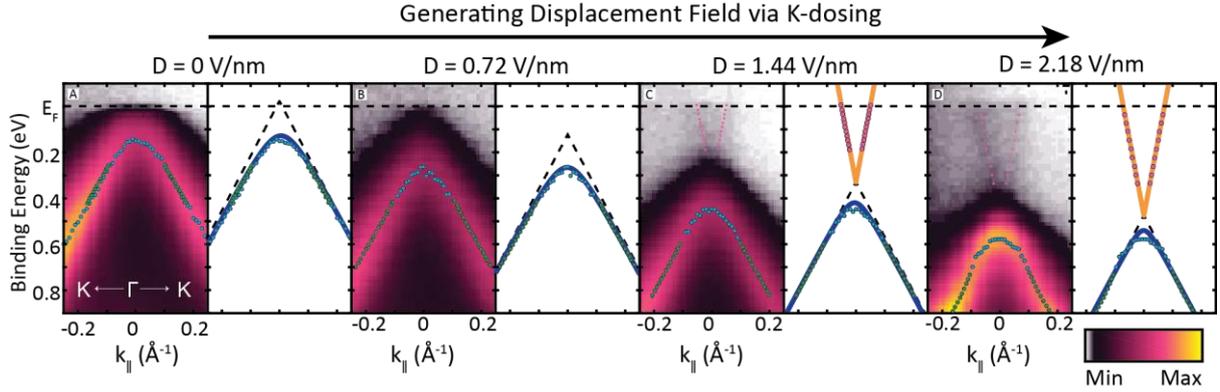

**Figure 3. Bandstructure modulation in ARPES in ML and BL Na₃Bi using K dosing.** (a)-(d) ARPES intensity plots showing the evolution of the band dispersion with K dosing, where the left panel represents the ARPES spectra and the right panel is the extracted MDC and EDC curves along with a bi-partite model fit (see SM5 for details). (a) Before in-situ K dosing. The hole band is located ~140 meV below $E_F$; (b) K dosing equivalent to a 0.72 Vnm$^{-1}$ displacement field; the hole band is now 257 meV below $E_F$; (c) K dosing equivalent to a 1.44 Vnm$^{-1}$ displacement field has n-type doped the system to an extent that an electron band has now emerged, separated from the hole band by ~100 meV; (d) K dosing equivalent to 2.18 Vnm$^{-1}$ displacement field results in further n-type doping with the band separation 90 meV.

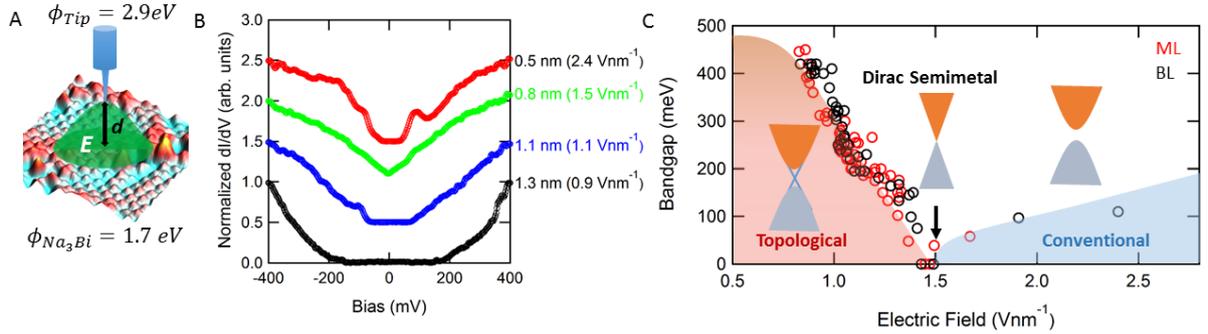

**Figure 4. Electric-field induced bandgap modulation of ML and BL Na₃Bi in STM.** (a) Schematic representation of a metallic tip (with work function $\phi_{Tip}$) at a fixed distance above the surface of Na₃Bi (with work function $\phi_{Na3Bi}$), with the difference in work function generating a localized electric field. The potential difference of ~1.2 eV is much larger than the bias applied for d$I$/d$V$ measurements. (b) Individual d$I$/d$V$ spectra taken on BL Na₃Bi at different tip-sample separations (electric fields) of: 1.3 nm (0.9 Vnm$^{-1}$) (black curve), 1.1 nm (1.1 Vnm$^{-1}$) (blue curve), 0.8 nm (1.5 Vnm$^{-1}$) (green curve) and 0.5 nm (2.4 Vnm$^{-1}$) (red curve). The spectra have been normalized and offset for clarity. (c) Bandgap extracted from d$I$/d$V$ spectra as a function of electric field for monolayer (red circles) and bilayer (black circles). At a critical field of ~1.5 Vnm$^{-1}$ the system is no longer gapped, above this a bandgap reopens in the conventional regime. The red and blue shaded regions represent guides to the eye.

SUPPLEMENTARY MATERIALS

Electric Field-Tuned Topological Phase Transition in Ultra-Thin Na$_3$Bi – Towards a Topological Transistor

James L. Collins[1,2,3], Anton Tadich[3,4], Weikang Wu[5], Lidia C. Gomes[6,7], Joao N. B. Rodrigues[6,8], Chang Liu[1,2,3], Jack Hellerstedt[1,2], Hyejin Ryu[9], Shujie Tang[9], Sung-Kwan Mo[9], Shaffique Adam[6,10], Shengyuan A. Yang[5], Michael. S. Fuhrer[1,2,3], Mark T. Edmonds[1,2,3*]

11. School of Physics and Astronomy, Monash University, Clayton VIC 3800, Australia
12. Monash Centre for Atomically Thin Materials, Monash University, Clayton VIC 3800 Australia
13. Centre for Future Low Energy Electronics Technologies, Monash University, Clayton VIC 3800 Australia
14. Australian Synchrotron, 800 Blackburn Road, Clayton, VIC 3168, Australia
15. Research Laboratory for Quantum Materials & EPD Pillar, Singapore University of Technology and Design, Singapore 487372, Singapore
16. Department of Physics and Centre for Advanced 2D Materials, National University of Singapore, 117551, Singapore
17. National Center for Supercomputing Applications, University of Illinois at Urbana-Champaign
18. Institute for Condensed Matter Theory and Department of Physics, University of Illinois at Urbana-Champaign
19. Advanced Light Source, Lawrence Berkeley National Laboratory, Berkeley, CA 94720, USA
20. Yale-NUS College, 6 College Avenue East, 138614, Singapore

**\* Corresponding author - <mark.edmonds@monash.edu>**

**S1. RHEED, LEED and XPS characterization of few-layer Na3Bi**
**S2. Band dispersion in $k_z$ from Photon energy-dependent ARPES**
**S3. Bandgap extraction from scanning tunneling spectra**
**S4. DFT Calculations for Na(2) vacancy Na3Bi layers and Z2 calculations**
**S5. ARPES band dispersion fitting and displacement field calculations**
**S6. dI/dV normalization and calculating tip-sample separation and electric field**

## S1. RHEED, LEED and XPS characterization of few-layer Na3Bi

In order to confirm that the few-layer films of Na$_3$Bi are high quality and epitaxial over large areas we performed reflection high-energy electron diffraction (RHEED), low energy electron diffraction (LEED) and X-ray photoelectron spectroscopy (XPS) as shown in Fig. S1. Figure S1(a) shows the characteristic RHEED pattern for Si(111) 7 x7 reconstruction along $\bar{\Gamma} - \bar{M}$, whilst Fig. S1(b) represents the RHEED pattern for few-layer Na$_3$Bi along $\bar{\Gamma} - \bar{K}$, consistent with RHEED reported on films of 15 unit cell thickness, where the lattice orientation of Na$_3$Bi is rotated 30º with respect to the Si(111) substrate [1]. Figure S1(c) shows the 1x1 LEED pattern consistent with growth of Na$_3$Bi in the (001) direction. The sharpness of the spots and absence of rotational domains indicates high-quality single crystal few-layer Na$_3$Bi over large area. Figure S1(d) shows the Bi 5d and Na 2p core levels of a few-layer Na$_3$Bi film taken at $hv$ = 100 eV, with the peak positions consistent with published results on 20 nm film and bulk Na$_3$Bi [1-3].

To rule out reaction of Na$_3$Bi with the Si substrate, we performed depth dependent XPS (by varying the photon energy in order to increase the kinetic energy of emitted photoelectrons, as a result increasing the mean free path) to examine the Na$_3$Bi and Si(111) interface. The Na and Bi core levels exhibited no additional components (data not shown). Figure S1(e) shows XPS of the Si 2p core level (reflecting the substrate) at hv = 350 eV (left panel) and hv = 850 eV (right panel). In each panel the black curve represents the Si 2p core level of the bare Si(111) 7 x7 and the red curve represents the Si 2p core level with few-layer Na$_3$Bi grown on top. In each case, the spectra have been normalized to the maximum in intensity and energy-corrected (to account for the small interfacial charge transfer that occurs) in order to overlay the core levels. The spectra have been offset for clarity. It is clear there is negligible change to the Si 2p core level after Na$_3$Bi growth, with no additional components or significant broadening arising, verifying that Na$_3$Bi is free-standing on Si(111).

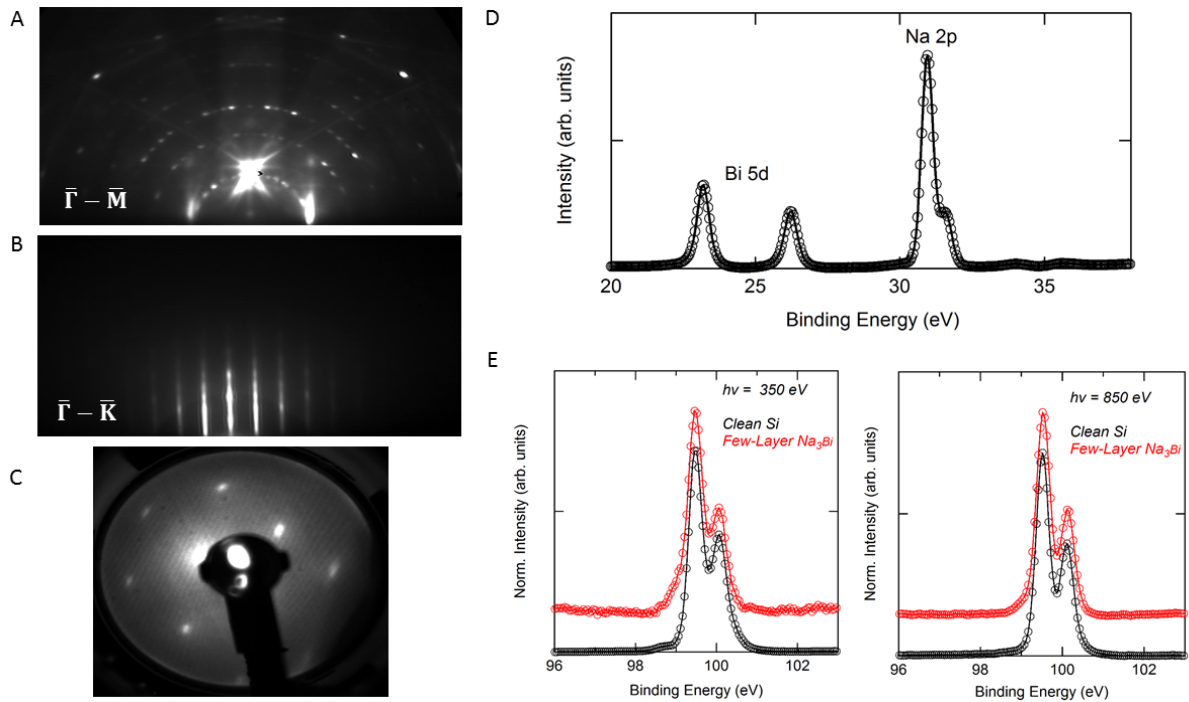

Figure S1. RHEED, LEED and XPS characterization of few-layer Na$_3$Bi. RHEED patterns of (a) Si(111) 7x7 reconstruction along $\bar{\Gamma} - \bar{M}$ direction and (b) few-layer Na$_3$Bi along the $\bar{\Gamma} - \bar{K}$ direction. (c) 1x1 LEED image of few-layer Na$_3$Bi taken at 32 eV. (d) XPS of Na 2p and Bi 5d core level taken at hv = 100 eV for few-layer Na$_3$Bi. (e) Normalized XPS of Si 2p core level taken at 350 eV (left panel) and 850 eV (right panel). Each panel shows the Si2p of the clean Si substrate (black curve) and with few-layer Na$_3$Bi grown on top. The spectra have been offset in intensity for clarity.

## S2. Band dispersion in $k_z$ from Photon energy-dependent ARPES

Photon energy-dependent ARPES can be utilized to determine whether a material possesses a 3D band dispersion, i.e. the binding energy $E_B$ depends not only on in-plane wavevectors $k_x$ and $k_y$, but also on out-of-plane wavevector $k_z$). To determine the momentum perpendicular to the surface requires measuring energy distribution curves as a function of the photon-energy in order to measure $E_B$ vs $k_z$, using the nearly free-electron final state approximation [4,5]:

$$k_z = \sqrt{\frac{2m}{\hbar^2}(E_k + V_0 - E_k \sin^2\theta)} \qquad (2.1)$$

where $\theta$ is the emission angle, m is the effective mass of electrons, $V_0$ is the inner potential (reflecting the energy difference between the bottom of the valence band to the vacuum level) and $E_k$ is the kinetic energy of the emitted photoelectrons where $E_k = h\nu - \Phi - E_B$ with $h\nu$ the photon energy, $\Phi$ the work function and $E_B$ the energy. At normal emission (i.e. $\theta = 0$) equation (2.1) simplifies to

$$k_z = \sqrt{\frac{2m}{\hbar^2}(E_k + V_0)} \qquad (2.2)$$

Therefore, using (2.2) and measuring energy distribution curves at normal emission as a function of photon energy we can directly measure $E_B$ vs $k_z$ assuming an inner potential, $V_0 = 12.5$ eV for Na$_3$Bi determined in [2].

Figure S2, shows a colour plot of $k_z$ as a function of binding energy (and reflects energy distribution curves taken at normal emission for photon energies between 45-55 eV. A flat band is observed near 0 eV (the Fermi energy) and represents the valence band maximum or Dirac hole band. This band possesses no dispersion in $k_z$ (i.e. no bulk band dispersion), verifying that few-layer Na$_3$Bi is indeed electronically 2D, unlike its thin-film and bulk counterparts.

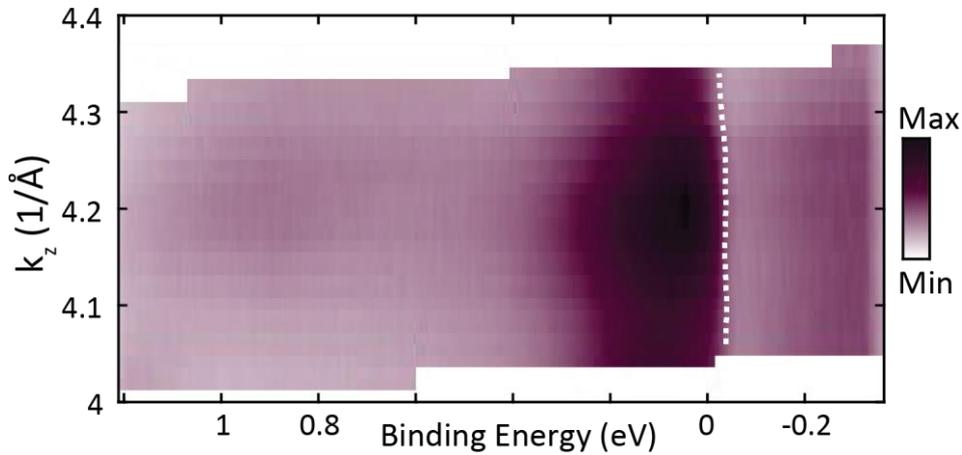

Figure S2. $k_z$ band dispersion of few-layer Na$_3$Bi from photon-energy dependent ARPES measurements. Photon-energy dependent ARPES of ML/BL Na$_3$Bi demonstrating effectively 2D dispersion. Cut through the photon-energy (hv=45-55 eV) dependent Fermi surface showing non-dispersion of the gapped Dirac valence band along $k_z$.

## S3. Bandgap extraction from scanning tunneling spectra

Determining the bulk electronic bandgap of mono- and bi-layer $Na_3Bi$ was achieved by performing scanning tunnelling spectroscopy ($dI/dV$ as a function of sample bias $V$) more than 5 nm away from step edges. The valence and conduction band edges in the local density of states (LDOS) are defined as the onset of differential conductance intensity above the noise floor. Due to the large variation in $dI/dV$ signal near a band edge, it was necessary to plot the logarithm of the $dI/dV$ curves for accurate band gap determination as shown in Figure S3 (a) and (b).

For measurements involving tuning the electric field by varying the tip-sample distance $dI/dV$ spectra were taken over a wide range of tunnelling currents (0.01-1nA), resulting in large changes in signal at band edges and a change in the relative magnitude of signal to noise. In order to unambiguously determine the magnitude of the gap without reference to the noise magnitude, we adopted the following procedure. Spectra were normalized to a relatively featureless point in the LDOS away from the band edge onset. The $dI/dV$ signal corresponding to a bias of -400 meV was chosen for normalization, as at large positive bias (400-600 meV) a spatially varying resonance feature was present; see Figure 2(e) of main manuscript. This resonance is most likely due to Na surface vacancies.[6] After the normalization procedure was completed for all spectra, we take the band edges as the point at which the $dI/dV$ has fallen to 0.1 of the normalized value. We find that this definition closely corresponds to the extrapolation of the conductance to zero on a linear scale (i.e. onset of conductance), and estimate that the error in determining the gap magnitude is ±10 meV.

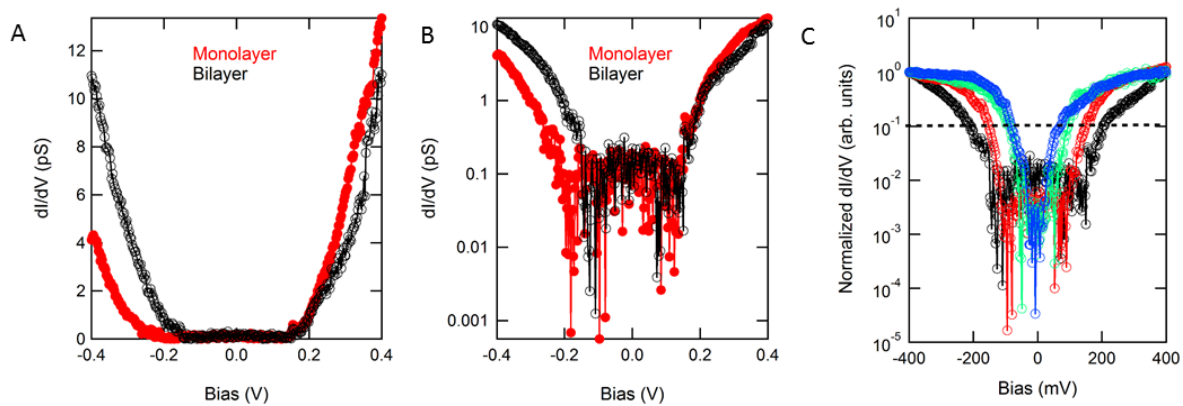

Figure S3. $dI/dV$ spectra taken for mono- (red) and bilayer (black) $Na_3Bi$ plotted on a (a) linear and (b) logarithmic scale. The logarithmic scale better accounts for the large change in intensity near the band edge. (c) Normalized $dI/dV$ spectra at various tip-sample separations illustrating that the onset in intensity typically occurs at a normalized $dI/dV$ signal of 0.1 within an error of ±10 meV.

## S4. DFT Calculations for Na(2) vacancy Na$_3$Bi layers and Z2 calculations

The calculated band structures of Na$_3$Bi monolayer and bilayer with Na(2) vacancy including spin-orbital coupling (SOC) are displayed in Fig. S4(b) and (e) respectively. Bandgaps of 0.30 eV (0.28 eV) and 0.22 eV (0.16 eV) are obtained for monolayer (without structural relaxation) and bilayer respectively, with the band gaps obtained from DFT using the Quantum Espresso code and the value obtained from the VASP package in brackets. The edge state spectrum is shown in Fig. S4(c) for monolayer and (f) for bilayer. The projected 1D Brillouin zone is shown in (a). One can clearly observe a Kramers pair of topological edge states, suggesting both mono- and bilayer Na$_3$Bi possess a nontrivial topological character. To prove these systems are non-trivial we determine the topological invariant of both systems. This is done by employing the Wilson loop method [7,8], in which one traces the evolution of the Wannier function centers, as plotted in Fig. 1(d) and 1(g). From the calculation, we confirm that both the monolayer and the bilayer Na$_3$Bi with Na(2) vacancy are topologically nontrivial with the invariant $\mathbb{Z}_2 = 1$

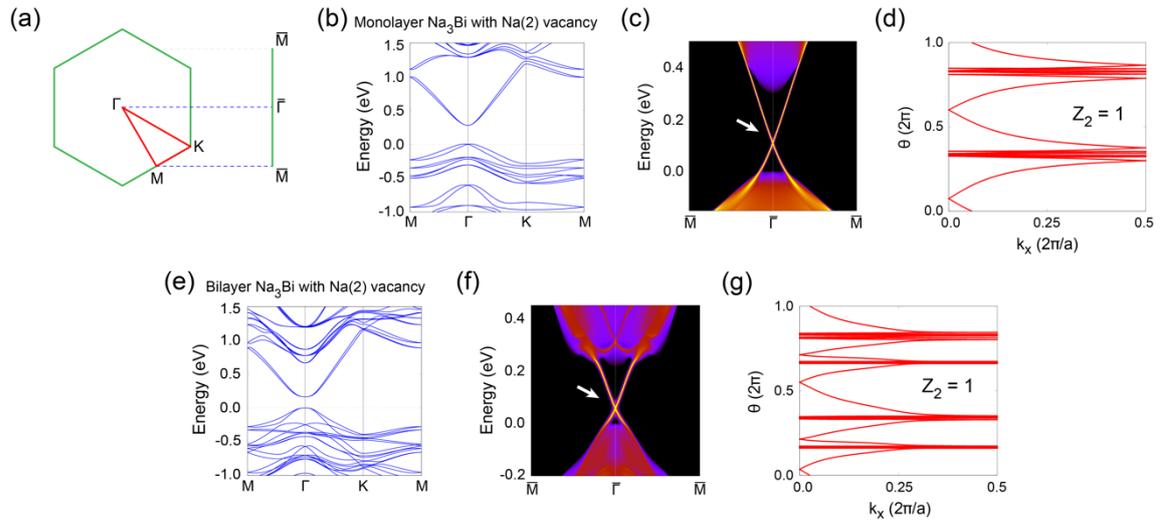

.
Figure S4 (a) Two-dimensional Brillouin zone for Na$_3$Bi layered structures. Here we also show the projected 1D Brillouin zone used for studying the edge spectrum. (b-g) Results for monolayer (b-f) and bilayer (e-g) Na$_3$Bi with Na(2) vacancies [with one Na(2) vacancy in a 2x2 supercell]. (b,e) Electronic band structures, where the energy zero is set to be at the valence band maximum. (c,f) Projected edge spectrum (edge along [010] direction), where pairs of $\mathbb{Z}_2$ topological edge states can be observed in the energy gaps (marked by the white arrows). (d,g) Calculated Wannier function center evolutions, which indicate a nontrivial $\mathbb{Z}_2$ invariant ($\mathbb{Z}_2 = 1$) for the bulk band structure in both monolayer and bilayer.

## S5. ARPES band dispersion fitting and displacement field calculations

### 5.1 Extracting and fitting the ARPES band dispersion of few-layer Na$_3$Bi:

Energy Dispersion Curves (EDC's) and Momentum Dispersion Curves (MDC's) are slices through constant-momentum and constant-energy of the photoemission spectra along high-symmetry directions (M - Γ - M) or (K - Γ - K) respectively. Band energy and momentum coordinates are extracted by Gaussian fitting of the photoemission intensity on a flat background (as shown in Fig. S5(b) and (c) by the blue circles). We find that band edges are extracted more reliably from EDCs, whilst MDCs peak positions are used at larger binding energies where clearly distinct peaks can be resolved (see left panel of Fig. S5(b)).

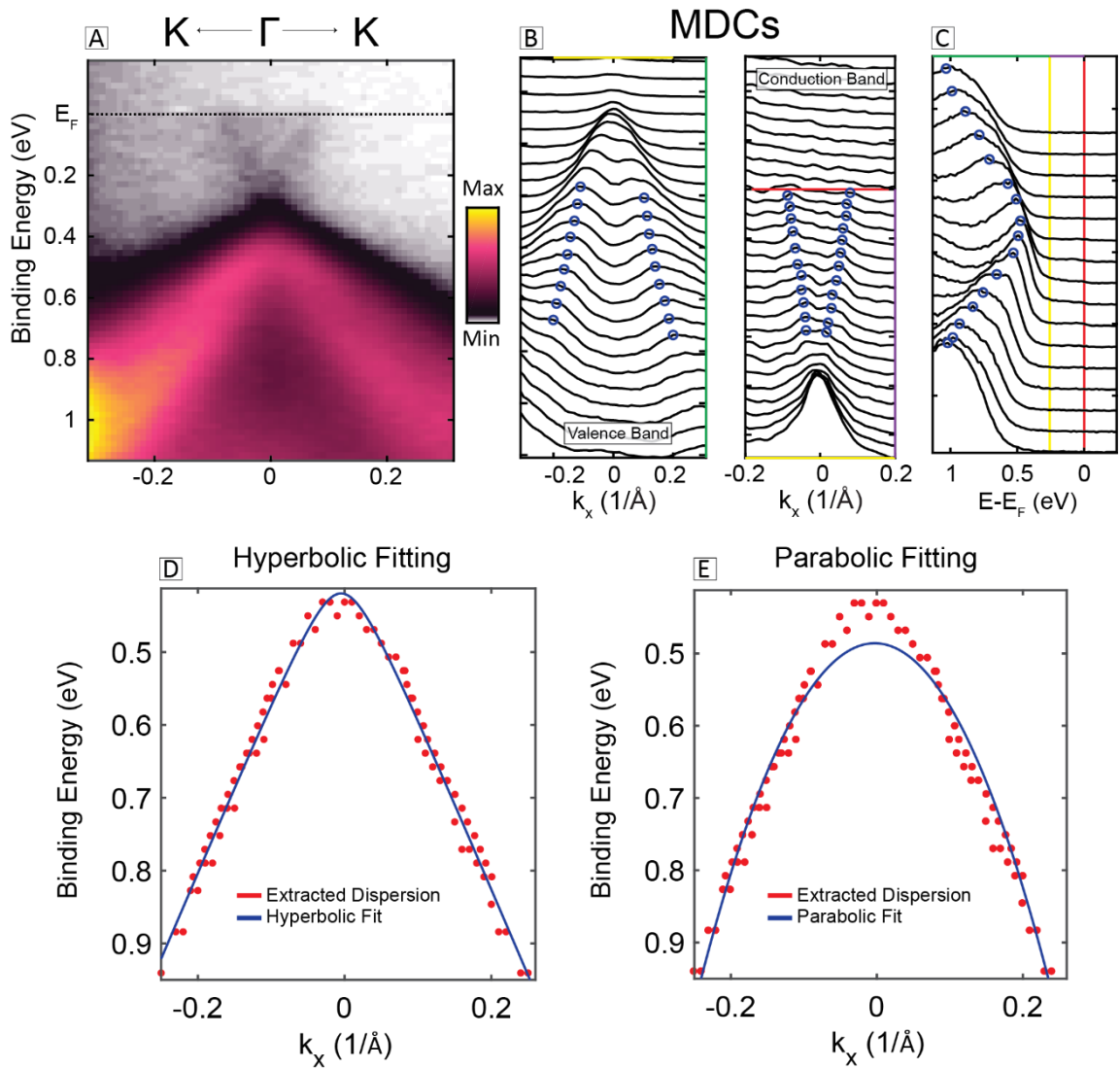

Figure S5a. Extracting and fitting the dispersion MDC and EDC points extracted from ARPES spectra on few-layer Na$_3$Bi. (a) ARPES intensity plot along K - Γ - K direction after 30 minutes of K-dosing. (b) MDC stack plots of the valence band (left panel) and conduction band (right panel) extracted from (a). (c) EDCs extracted from (a). (d,e) Fitting extracted band coordinates by MDC and EDC analysis (red) to a hyperbola (as in d) and parabola (as in e), showing that Na$_3$Bi Dirac-like bands are best matched with a hyperbola function.

The measured bands are observed to fit hyperbolae as shown in Figure S5(d), as expected of Dirac systems [2]. Fits of the bands to a parabola are much poorer, as shown in Fig. S5(e). In order to accurately model the band dispersion of a 2D gapped Dirac system we use a bi-partite model for the valence (p) and conduction (n) bands that assumes the form

$$(E_{B,i} - D)^2 = \Delta_i^2 + \hbar^2 v_{F,i}^2 (k + k_0)^2, i \in p, n \quad (5.1)$$

where $\Delta$ represents an energy-gap corresponding to the semi-major axes, $v_F$ the band Fermi-velocity, and $D$ a doping or energy-shift of the bands.

A near-constant $v_F$ is measured for the valence and conduction bands for both films, with near-isotropic dispersion in $k_x, k_y$ (also shown in S6), and is not seen to change with K-dosing. Hence we take $v_{F,n}$ and $v_{F,p}$ to be a global fit parameter, with best fit values $v_{F,n} \approx 1 \times 10^6 m/s$; $v_{F,p} \approx 3 \times 10^5 m/s$.

We then fit the valence band photoemission i.e. the negative solution for (5.1) using the global $v_{F,p}$ parameter, allowing us to determine $\Delta_p$ and $D$ as a function of K-dosing. A monotonic increase of $D$ with K-dosing is observed as expected, reflecting the shift of the valence band to larger binding energy.

The photoemission intensity of the electron band is two orders of magnitude less than that of the valence band – possibly due to the different orbital characters of the two bands resulting in a lower intensity due to matrix element effects. Due to the large band gap of few-layer Na$_3$Bi, the conduction band lies well above the Fermi level in the as-grown film, meaning that significant charge transfer from K-dosing is needed to n-type dope the film in order to observe the conduction band. As such the fitting parameter $\Delta_n$ for the electron bands can only be determined once the conduction band is resolvable below E$_F$, and in addition further seen to match with the valence band determined value for $D$. Values for $\Delta_{n,p}$ as a function of electric displacement field are addressed in **S5.3**

### 5.2 *Mapping from K-deposition to space-charge generated Electric Displacement Field*:
Potassium deposited on the Na$_3$Bi surface donates electrons leaving a positive K$^+$ ion behind, producing a uniform planar charge density. This is equivalent to a parallel plate capacitor, allowing the electric displacement field to be determined across the Na$_3$Bi film.

The charge transfer to the system cannot be directly inferred when a Fermi surface cannot be clearly resolved, so our calculations make use of the conduction band Fermi surface that becomes distinct after 15 minutes of K-dosing. As seen in Figure S6(a), the n-type Fermi surface is a nearly isotropic Dirac cone. By measuring $k_F$ as a function of K-dosing either from EDCs or a Fermi surface map as in Fig. S6(b), the charge density can be directly calculated using

$$n(k_F) = \frac{g}{4\pi} k_F^2, \quad (5.2)$$

where a band degeneracy of $g = 4$ can be taken for Dirac systems [6]. The charge density $n(k_F)$ is also consistent with the assumption of a Dirac dispersion centred at $D$, i.e. $k_F = D/\hbar v_F$.

The change in $n(k_F)$ as a function of K-dosing is approximately 2 x 10$^{12}$ cm$^{-2}$ between consecutive K-dosing until the 50-minute mark (where charge saturation occurs). By assuming that in this regime every K atoms donates one electron and a constant dose rate we can extrapolate the total $n(k_F)$ back to the doping of the as-grown film growth. For the as-grown film this corresponds to a p-type doping of 4x10$^{12}$

cm$^{-2}$. This value is reasonable given that the bulk band gap measured with STS is >400 meV and the separation from the valence band edge to the Fermi level from the ARPES data is 140 meV. From this as-grown doping we can then calculate the electric displacement field from Gauss' law, shown in Figure S6(c).

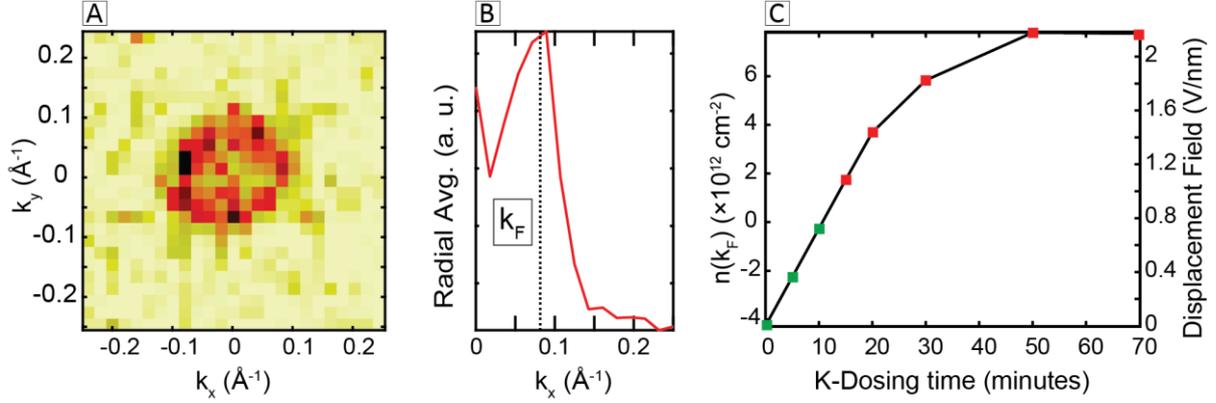

S6 (a) Electron-band Fermi-surface of few-layer Na$_3$Bi after 30 minutes of K-dosing. (b) Radially averaged momentum profile through the Fermi surface, showing the ring structure at $k_F$. (c) Calculated charge-density assuming degeneracy of bands $g = 4$ for using (5.2) and the corresponding electric displacement field associated with the associated net charge transfer from the undoped film as a function of K-dosing time. Red points are as-measured, and green points are extrapolated based on the $E_F$ shifting rate with K-dosing between 15-50 minutes.

**5.3 *Electric displacement field dependence of few-layer Na$_3$Bi bandstructure from ARPES*:**

Next we map the calculated $\Delta_{n,p}$ (which reflects the size of the bandgap) at different K-dosing to a corresponding electric displacement field as shown in Figure S7(a). The purple circles in (a) represent where $\Delta_{n,p}$ are directly extracted from the bi-partite model to the experimental data. At low displacement field, where the conduction band is still above the Fermi level, we cannot directly obtain a value for $\Delta_n$. As such we estimate the size of $\Delta_n$ at these low displacement fields using the ratio $\frac{\Delta_p + \Delta_n}{\Delta_p} \approx 1.4$, which is directly obtained from the purple points. For the as-grown sample this yields a value of ~320 meV, which given we cannot directly measure the conduction band edge is in reasonable agreement with the experimental result from STS, and the theoretical DFT value. The reduction in bandgap is consistent with the independently measured gap-closing from STS in Fig. 4(c) of the main manuscript, with the relative energy separation of the electron and valence bands narrowing monotonically with increasing field. When the sum of $\Delta_{n,p}'s$ approaches 100 meV (corresponding to displacement fields >1.5 Vnm$^{-1}$), we are approaching the intrinsic energy broadening of the bands (particularly the valence band). Therefore, we cannot definitely conclude from ARPES measurements whether the band gap completely closes after this point or is re-opened again. However, it is worth noting these fields are very close to the field required to close the bandgap in STS measurements.

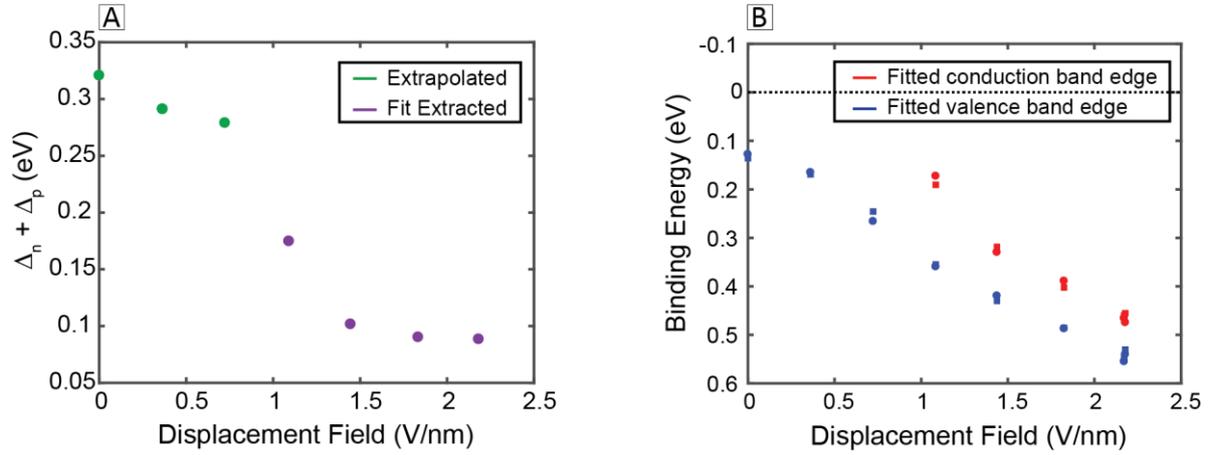

Figure S7. Electric displacement field dependence of topological insulator ML/BL Na$_3$Bi bands near $E_F$. (a) The sum of Δ parameters from the best fit of (3) to ARPES dispersion vs. applied electric displacement field. Both $\Delta_{n,p}$ are directly calculated from the high-field (purple) measurements, however at low-field (green) measurements, $E_F$ is not sufficiently shifted for the electron dispersion to be clearly resolved. Here we have used the ratio $\frac{\Delta_p+\Delta_n}{\Delta_p} \approx 1.4$ measured from the purple points to extrapolate $\Delta_n$ undosed film. (b) The valence edge (blue) and conduction edge (red) calculated from the fitted ARPES dispersion is shown for two high-symmetry directions: circles and squares as K - Γ - K and M - Γ - M respectively.

## S6. Calculating tip-sample separation and electric field

In scanning tunnelling microscopy the tunnelling current follows

$$I = A \times exp\left(\frac{-2\sqrt{2m\Phi}}{\hbar}(z - z_0)\right) \qquad (6.1)$$

where $\Phi$ is the work function of the energy barrier and $z$ and $z_0$ the tip and sample positions respectively. This allows the work function of the barrier to be obtained by measuring the tunnelling current as a function of tip position, then extracting the slope of $\ln(I)/z$:

$$\frac{d\ln(I)}{dz} = \frac{-2\sqrt{2m\Phi}}{\hbar}$$

Figure S6 shows logarithmic plots of the tunnelling current as a function of relative distance for (a) Au(111) (Bias 500 mV and 100 pA) and (b) thin film $Na_3Bi$ (Bias 300 mV and 100 pA). For Au(111) the characteristic linear dependence is observed with current increasing from 0.01 nA to 10nA by moving the tip 3Å closer to the surface. However, it is immediately clear that very different behaviour occurs for $Na_3Bi$ in Fig. S6(b). At low tunnelling current a linear dependence with distance is observed, but as the distance from tip and sample decreases the current saturates to a value around 1nA, which occurs over a length scale of ~1 nm. This corresponds to the barrier height going to zero as the tip approaches the sample surface. This behaviour occurs in low work function materials and is a result of two mechanisms. First, the effective barrier height (defined from the slope of $d\ln(I)/dz$) is often significantly lower than the actual barrier height, and gets lower the closer to tip gets to the sample surface [9]. The second effect is even more pronounced and results in a lowering of the energy barrier due to the image potential, which acts to increase the current at large distances by a fixed amount such that the apparent barrier height at large distances is the work function, but the tip-sample distance is larger than expected. This increased distance due to the image potential may be as large as ~15 Å/[$\Phi$(eV)] [10].

To understand and quantify the tip-sample distance in $Na_3Bi$, we first extract the slope from the linear region in Fig. S6(b) (as shown by the black line). This fit yields a slope of 1.6 Å$^{-1}$ and A=1.93x10$^{-10}$A. Accounting for the bias of 300 mV yields a work function or barrier height of 2.3 eV. $I$ vs $Z$ measurements were taken at various negative and positive bias values and yielded work functions of 2.3 ± 0.05 eV. This confirms a very low barrier height, meaning the effect of the mirror potential will lead to an increased distance of 6.5 Å.

From the fit to Fig. S6(b) and extracted values of the slope of 1.6 and A=1.96x10$^{-10}$, we can determine the distance that corresponds to point contact. Point contact corresponds to a conductance on order the conductance quantum, approximately 40 µS, so converting our current data into conductance we are able to calculate that contact occurs at $Z = z_0 = $ -6.7 Å in Fig. S6(b).

We are now in a position to estimate the tip-sample separation from the $dI/dV$ measurements. For each $dI/dV$ curve we adopt the following procedure:
   (1) Extract a relative Z distance for each $dI/dV$ measurement by referencing the tunnelling current at -400 mV (the starting bias for each measurement) to the $I$ vs. $Z$ data in Fig. S6(b).
   (2) Add the point contact value of 6.7 Å and mirror potential offset of 6.5 Å to this value.

The four $dI/dV$ spectra in Fig. 4(b) of the main manuscript have the following:

***Black curve:*** Tunnelling current of 100 pA corresponds to relative z distance of 0 Å. Adding 13.2 Å to account for point contact and mirror potential yields a tip-sample separation of 1.32 nm.
***Blue curve:*** Tunnelling current of 440 pA corresponds to relative z distance of -2.1 Å. Adding 13.2 Å to account for point contact and mirror potential yields a tip-sample separation of 1.1 nm.
***Green curve:*** Tunnelling current of 800 pA corresponds to relative z distance of -4.9 Å. Adding 13.2 Å to account for point contact and mirror potential yields a tip-sample separation of 0.84 nm.
***Red curve:*** Tunnelling current of 920 pA corresponds to relative z distance of -8.3 Å. Adding 13.2 Å to account for point contact and mirror potential yields a tip-sample separation of 0.5 nm.

We now have an estimate of the tip-sample separation, so to calculate a displacement field for each of our d$I$/d$V$ measurements we need to calculate the potential difference i.e. the work function difference between the metallic tip and the few-layer Na$_3$Bi. The measured barrier height of 2.3 eV is an average of the tip and sample work functions, $\Phi_{Barrier} = (\Phi_{Tip} + \Phi_{Na_3Bi})/2$, meaning we need to know either the tip work function or measure the work function of the Na$_3$Bi.

The Kelvin probe technique was utilized to measure the work function of few-layer Na$_3$Bi. The work function was determined by measuring the contact potential difference of the Na$_3$Bi relative to a gold reference of known work function (determined by photoelectron spectroscopy secondary electron cutoff measurements). A work function for few-layer Na$_3$Bi of 1.7 ± 0.05 eV was measured using this technique. This value and the 2.3 eV potential barrier gives a tip work function of 2.9 eV, and a potential difference between tip and sample of 1.2 eV. It should be noted that this is significantly lower than the expected value for a PtIr tip, suggesting that Na atoms that have been picked up by the tip and consequently lowering the work function (work function of Na is 2.23 eV). Whilst Na atoms were picked up they have little influence on the tip density of states, as spectroscopy performed on Au(111) after measurements on Na$_3$Bi were completed revealed a flat LDOS near the Fermi energy.

Using the calculated potential difference and tip-sample separation from above allows us to calculate electric fields for d$I$/d$V$ spectra on ML and BL Na$_3$Bi as shown in Fig. 4(b) for BL and Fig. S7 for ML, with the full data set of bandgap as a function of electric field shown in Fig. 4(c).

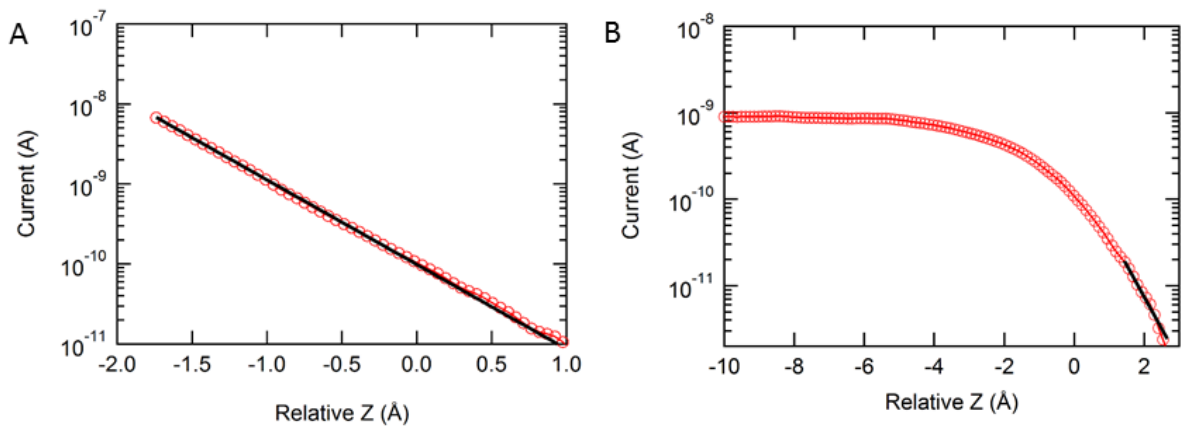

Figure S6. Tunnelling current as a function of relative distance for (a) Au(111) and (b) thin-film Na$_3$Bi. The black line represents a linear fit, in order to extract the tip-sample barrier height.

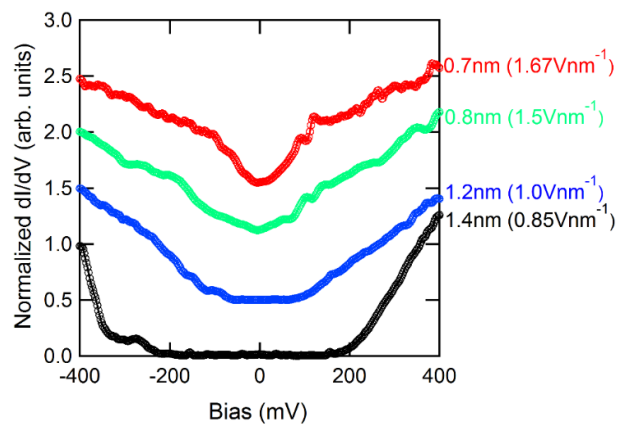

**Figure S7.** Individual d$I$/d$V$ spectra taken on ML Na$_3$Bi at different tip-sample separations (electric field) of: 1.4 nm (0.85 Vnm$^{-1}$) (black curve), 1.2 nm (1.0 Vnm$^{-1}$) (blue curve), 0.8 nm (1.5 Vnm$^{-1}$) (green curve) and 0.7 nm (1.67 Vnm$^{-1}$) (red curve). The spectra have been normalized and offset for clarity.